\documentclass[12pt]{article}
\usepackage{amsmath,amssymb,epsfig,amsfonts,epsfig,graphicx}
\usepackage{natbib}
\usepackage{epsfig}
\usepackage{epstopdf}

\usepackage{amssymb}
\usepackage{amsmath} \usepackage{stmaryrd}
\usepackage{epic} \usepackage{eepic}

\newcommand{\beq}{\begin{equation}}
\newcommand{\eeq}{\end{equation}}
\newcommand{\bea}{\begin{eqnarray}}
\newcommand{\eea}{\end{eqnarray}}
\newcommand{\beas}{\begin{eqnarray*}}
\newcommand{\eeas}{\end{eqnarray*}}

\renewcommand{\rho}{\varrho}
\renewcommand{\phi}{\varphi}
\newcommand{\nn}{\nonumber}

\newcommand{\mbs}[1]{\mathbf{#1}}

\def\bA{{\mbs{A}}} \def\bB{{\mbs{B}}} \def\bC{{\mbs{C}}}
\def\bD{{\mbs{D}}}  
  \def\bI{{\mbs{I}}}
  
\def\bM{{\mbs{M}}} \def\bN{{\mbs{N}}}

\def\bg{{\mbs{g}}} \def\bh{{\mbs{h}}}

\newcommand{\Z}{\mathbb{Z}}
\newcommand{\R}{\mathbb{R}}

\newcommand{\trace}{\operatorname{Tr}}
\newcommand{\I}{\mathbf{I}}
\newcommand{\II}{\mathbf{II}}
\newcommand{\Id}{\mathbf{Id}}
\newcommand{\E}{\mathbf{E}}

\newcommand{\D}{\mathbf{D}}
\newcommand{\lat}{\mathbf{a}}
\newcommand{\latt}{\mathbf{b}}
\newcommand{\h}{\mathbf{h}}

\usepackage [colorlinks =true, hyperindex =true]{ hyperref}

\begin{document}

\begin{center}
{\bf \large Universal formulae for the limiting elastic energy \\ of membrane networks} \\[0.5cm]  

Bernd Schmidt \\[0.1cm] 
{\footnotesize Institut f{\"u}r Mathematik, Universit{\"a}t Augsburg\\ 
Universit{\"a}tsstr.\ 14, 86159 Augsburg, Germany}\\[0.5cm]

Fernando Fraternali \\[0.1cm] 
{\footnotesize Department of Civil Engineering, University of Salerno,
84084 Fisciano(SA), Italy, and \\
Division of Engineering,
King's College London,
Strand, London  WC2R 2LS,
UK}
\end{center}

\begin{abstract}
We provide universal formulae for the limiting stretching and bending energies of triangulated membrane networks endowed with nearest neighbor bond potentials and cosine-type dihedral angle potentials. The given formulae account for finite elasticity and solve some deficiencies of earlier results for Helfrich-type bending energies, due to shape-dependence and sensitivity to mesh distortion effects of the limiting elastic coefficients. We also provide the entire set of the elastic coefficients characterizing the limiting response of the examined networks, accounting for full bending-stretching coupling. We illustrate the effectiveness of the proposed formulae by way of example, on examining the special cases of cylindrical and spherical networks covered with equilateral triangles, and discussing possible strategies for the experimental characterization of selected elastic moduli. 
\end{abstract}

{\bf Key words:}
Membrane networks,  Dihedral angle potentials, Nearest neighbor bond potentials, Multiscale modeling, Limiting energies, Bending stiffness.

\section{Introduction}
Polyhedral membrane models endowed with bond and dihedral angle potentials (hereafter referred to as \textit{membrane networks}) are widely employed in many different physical and engineering problems, including  molecular dynamics (MD) simulations of bio- and nano-structures (refer, e.g., to the recent review papers \citeauthor{Muller:2006}, \citeyear{Muller:2006}, \citeauthor{Tu:2008}, \citeyear{Tu:2008} and references therein), and non-conforming models of plates and shells (\citeauthor{Davini:2000}, \citeyear{Davini:2000}, \citeauthor{Angelillo:2006} \citeyear{Angelillo:2006}), among others. Referring to the special case of the red blood cell membrane, the bond potentials reproduce the protein filaments forming the membrane skeleton, while the dihedral angle potentials mimic the bending rigidity endowed by the lipid bilayer  (\citeauthor{Tu:2008}, \citeyear{Tu:2008}; \citeauthor{Discher:1998}, \citeyear{Discher:1998}; \citeauthor{Marcelli:2005}, \citeyear{Marcelli:2005}; \citeauthor{Suresh:2006}, \citeyear{Suresh:2006}; \citeauthor{Hale:2009}, \citeyear{Hale:2009}; \citeauthor{Hartmann:2010}, \citeyear{Hartmann:2010}). In such a case, the discrete nature of the network is motivated by the finite size of the protein filaments, and the convenience of numerical approaches like MD and Monte Carlo simulations (\citeauthor{Kroll:1992}, \citeyear{Kroll:1992}).  On the other hand, the use of membrane networks for the development of \textit{lumped strain} models of plates and shells allows for capturing local effects due, e.g.,  to strain localization, plastic hinges and folding-like buckling (\citeauthor{Davini:2000}, \citeyear{Davini:2000}; \citeauthor{Angelillo:2006}, \citeyear{Angelillo:2006}).

A key question for such models is the determination of the continuum limit of stretching and bending energies, when the nearest neighbor distance $r$ becomes infinitesimally small (\citeauthor{Seung:1988}, \citeyear{Seung:1988}; \citeauthor{Gompper:1996}, \citeyear{Gompper:1996}; \citeauthor{Disher:1997}, \citeyear{Disher:1997}; \citeauthor{Zhou:1997}, \citeyear{Zhou:1997}; \citeauthor{Schmidt:2006}, \citeyear{Schmidt:2006}; \citeauthor{Schmidt:2008}, \citeyear{Schmidt:2008}; \citeauthor{Hartmann:2010}, \citeyear{Hartmann:2010}). While the limit form of the membrane energy is well established in the literature, at least for harmonic bonds (\citeauthor{Seung:1988}, \citeyear{Seung:1988}; \citeauthor{Disher:1997}, \citeyear{Disher:1997}; \citeauthor{Zhou:1997}, \citeyear{Zhou:1997}) some questions still remain open regarding the continuum limit of the bending energy, when cosine-type dihedral angle potentials are employed. As a matter of fact, earlier results in terms of Helfrich-type limiting energies  for triangulated networks suffer from deficiencies due to shape-dependence and mesh distortion effects (\citeauthor{Seung:1988}, \citeyear{Seung:1988}; \citeauthor{Gompper:1996}, \citeyear{Gompper:1996}; \citeauthor{Nelson:2004}, \citeyear{Nelson:2004}; \citeauthor{Espriu:1987}, \citeyear{Espriu:1987}; \citeauthor{Baillie:1990}, \citeyear{Baillie:1990}). Different is the case of bending energies based on nodal averages of dihedral angles, which exhibit a more uniform limit behavior (\citeauthor{Gompper:1996}, \citeyear{Gompper:1996}).

We deal in the present paper with the derivation of limiting stretching and bending energies of static membrane networks undergoing large in-plane and out-of-plane deformations. 
We consider a discrete mass spring system with nearest neighbor and dihedral spring interaction subject to deformation, and provide explicit formulae for the energy in the system when the number of particles (vertices) tends to infinity (i.e., $r \rightarrow 0$). The limiting energy only depends on the geometric data of the deformation mapping. As compared to previous works on the limiting bending energies of membrane networks (\citeauthor{Seung:1988}, \citeyear{Seung:1988}; \citeauthor{Gompper:1996}, \citeyear{Gompper:1996}), the present study follows a different from discrete to continuum approach, by interpreting the discrete deformation of the network as the restriction of a continuous deformation mapping $f$ to a triangular lattice. This is in agreement with the Cauchy-Born rule of crystalline matter (\citeauthor{Ericksen:2008}, \citeyear{Ericksen:2008}), and avoids mesh-dependent effects (\citeauthor{Muller:2006}, \citeyear{Muller:2006}; \citeauthor{Gompper:1996}, \citeyear{Gompper:1996}; \citeauthor{Nelson:2004}, \citeyear{Nelson:2004}). In the bending-dominant deformation regime, when the deformed configuration of the network approximatively coincides with an equilateral triangulation of a curved surface, the formula we obtain for the limiting bending energy reduces to 
\begin{align*}
E^{\rm bend}(f) 
= \frac{D}{4 \sqrt{3}} \int_U (3 H^2 - 8 K) \, du,
\end{align*}
where: $D$ is the dihedral constant; $U$ is the reference configuration: $H$ is twice the mean curvature, and $K$ is the Gaussian curvature. The formula above corresponds to a Helfrich-type energy $\int_U (\kappa_H H^2 / 2 +\kappa_G K) \, du$, with $\kappa_G = - 4 \kappa_H / 3$ and $\kappa_H = \sqrt{3} D / 2$ for any geometry, $\kappa_H$ and $\kappa_G$ being the bending and Gaussian rigidities, respectively (\citeauthor{Helfrich:1973}, \citeyear{Helfrich:1973}). This formula in particular applies to lipid membranes which show a fluid-like behavior in the in plane coordinates. Such a behavior leads to the relaxation of all stretching energy contributions and thus results in a pure bending energy functional. Those membranes can be described accurately within our framework of spring networks if we allow for a distribution of defects in the underlying lattice network on a mesoscopic scale. In this way one can find meshes with almost no stretching contributions in any deformed configuration. Also note that the energy then only depends on the curvature tensor of the surface $f(U)$ and $f$ is merely some parameterization of that surface used to express the bending energy $\frac{D}{4 \sqrt{3}} \int_{f(U)} (3 H^2 - 8 K) \, dS$ in local coordinates. 

For general deformations of triangulated networks with both stretching and bending contributions, however, the limiting energy expressions will naturally depend on the lattice reference configuration. As a matter of fact, the prediction of mesoscopic  
values of the referential moduli from MD simulations, eventually  
different from region to region of the model, allows one to adequately  
inform simulations at the continuum scale, such as, e.g., Lagrangian  
finite element models. Moreover, stability analyses are naturally  
performed through referential (or thermodynamic) elastic constants, as  
shown, e.g., in \citeauthor{Zhou:1996}, \citeyear{Zhou:1996}; \citeauthor{Zhou:1997}, \citeyear{Zhou:1997}; \citeauthor{Disher:1997}, \citeyear{Disher:1997}; \citeauthor{Hess1997}, \citeyear{Hess1997}.  

Concerning the present estimates of the bending and Gaussian rigidities, we wish to remark that \citeauthor{Seung:1988} (\citeyear{Seung:1988}) found the different result $\kappa_G = - \kappa_H$ and obtained a limiting bending energy of the form $\kappa_H/2 \int_U (H^2   - 2 K ) \, du$. The same authors also estimated  $\kappa_H=\sqrt{3}D/2$ by applying this formula to a cylindrical network. Subsequently, \citeauthor{Gompper:1996} (\citeyear{Gompper:1996}) made use of the result $\kappa_G = - \kappa_H$ to derive a different value of $\kappa_H$ for the sphere. These authors computed the limiting value of a discrete notion of the Laplacian part of the Helfrich energy ($\beta{\cal H}^{\rm Lap}=\int_U  H^2 / 2  \, du$), and subtracted the total curvature of the sphere $K^{\rm tot}=\int_U  K  \, du = 4 \pi$ to $\beta{\cal H}^{\rm Lap}$. On matching $\kappa_H (\beta{\cal H}^{\rm Lap} - K^{\rm tot})$ to the limiting value of the total dihedral energy of the sphere (${4 D\pi}/{\sqrt{3}}$, cf.\ formula (19) of \citeauthor{Gompper:1996}, \citeyear{Gompper:1996}), they finally obtained $\kappa_H=\sqrt{3}D/3$ for such a geometry.  

It is easily seen that the present limiting bending energy $\frac{D}{4 \sqrt{3}} \int_U (3 H^2 - 8 K) \, du$ exactly reduces to the limiting dihedral energies of both the sphere and the cylinder, with $\kappa_H=\sqrt{3}D/2$ in each case (cf.\ Sec.\ \ref{cases}). 
We wish to remark that the scale bridging approach behind the `universal' results $\kappa_G = - 4 \kappa_H / 3$ and $\kappa_H = \sqrt{3} D / 2$ assumes that the discretized surface matches the limiting shape in correspondence with the vertices of the triangulation. In more detail, such results are obtained looking at the discrete dihedral energy under the above Cauchy-Born type assumption, without making any `a priori' guess on the  $\kappa_G /  \kappa_H$ ratio. Interestingly, the present estimates of $\kappa_H$  and $\kappa_G$ correspond to those obtained in \citeauthor{Lidmar:2003}, \citeyear{Lidmar:2003} for the bending-dominant regime, through direct arguments.
These authors `assume' that the limiting bending energy can be written as a
Helffrich-type energy with two free parameters ($\kappa_H$  and $\kappa_G$, namely) in such a regime.
Then they use the two explicit cases of a sphere and a cylinder to determine the above parameters. 
An analogous result is deduced in \citeauthor{Fedosov:2009}, \citeyear{Fedosov:2009} for the special case of a spherical shell.
As we already observed, the present result for the bending energy is obtained without any a-priori assumptions on the structure of the limiting energy and the shape of the network and, as far as we know, 
represents the first complete proof to-date that the limiting continuum energy of membrane networks endowed with cosine-type dihedral angle potentials is actually a Helffrich type functional in the bending-dominant regime.

The general formula we provide for the bending energy accounts for the coupling of large stretching and bending deformations, which is a key feature of finite deformation shell theories (\citeauthor{Naghdi72}, \citeyear{Naghdi72}). It allows us to handle arbitrarily distorted triangulations, thus resolving the distortion sensitivity effects that derive from the use of a pure bending energy in the continuum limit (\citeauthor{Gompper:1996}, \citeyear{Gompper:1996}; \citeauthor{Nelson:2004}, \citeyear{Nelson:2004}; \citeauthor{Espriu:1987}, \citeyear{Espriu:1987}; \citeauthor{Baillie:1990}, \citeyear{Baillie:1990}). Concerning the limiting stretching energy, our findings confirm the results given in \citeauthor{Seung:1988}, \citeyear{Seung:1988}; \citeauthor{Disher:1997}, \citeyear{Disher:1997}; \citeauthor{Zhou:1997}, \citeyear{Zhou:1997}.

The paper is organized as follows. We first introduce the examined network model and the basic assumptions of the adopted scale bridging approach in Sec.\ \ref{model}. Next, we develop universal formulae for the limiting stretching and bending energies of such a network in Secs.\ \ref{stretching} and \ref{bending}, respectively. Elastic moduli are determined for the resulting effective theory in Sec.\ \ref{sec:moduli}. In Sec.\ \ref{cases}, we examine the special cases of cylindrical and spherical networks covered with equilateral triangles and undergoing pure bending. Sec.\ \ref{expvalidation} discusses a micropipette aspiration test as a nontrivial example for a general stretched and bent surface. We end by summarizing the main conclusions of the present work in Sec.\ \ref{conclusions}.

\section{The model}\label{model}

\begin{figure}[htbp]
  \begin{center}
    \epsfig{file=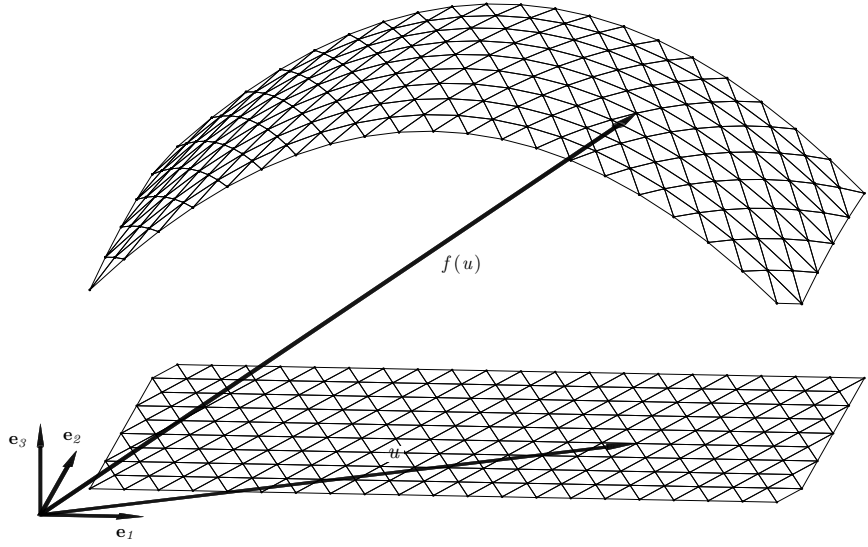,angle=0,width=80mm, trim = 120mm 50mm 120mm 50mm} 
   \caption{Reference and deformed configurations of a membrane network generated by a triangular lattice.}
    \label{fig:membrane_network}
  \end{center}
\end{figure}

Let ${\cal L}$ denote the triangular lattice ${\cal L} = \begin{footnotesize} \begin{pmatrix} 1 & 1/2 \\ 0 & \sqrt{3}/2 \end{pmatrix} \end{footnotesize} \Z^2$ in the Euclidean plane. We consider a rescaled piece ${\cal L}_r = r {\cal L} \cap U$ of ${\cal L}$, where $r > 0$ and $U \subset \R^2$ open, which we consider as our reference configuration. A mapping $f : U \to \R^3$ (which we assume to be sufficiently smooth up to the boundary) defines a discrete deformation when restricted to ${\cal L}_r$ (Fig. \ref{fig:membrane_network}): 
$$ {\cal L}_r \ni u \mapsto f(u). $$  

The energy of the system shall be due to stretching and bending contributions:
$$ E_r(f) = E^{\rm stretch}_r(f) + E^{\rm bend}_r(f). $$  
For large deformations we assume that the stretching energy is given by nearest neighbor interactions 
\begin{align}\label{eq:E-stretch}
E^{\rm stretch}_r(f) = \sum_{u,v \in {\cal L}_r \atop |u-v| = r} W_r(|f(u) - f(v)|) 
\end{align} 
with $W_r(s) = r^2 W(s/r)$ for some smooth $W$. We note that such a scaling assumption is necessary in order to account for uniform response to stretching on the nearest neighbor scale $|u-v| = r$ that is comparable to the size of an elementary lattice cell. For quadratic $W(s) = \frac{k}{2}(s-1)^2$ one has $W_r(s) = \frac{k}{2}(s - r)^2$. 
By way of contrast, in the regime of infinitesimal elasticity, where $(Df)^T Df - \Id = O(r)$, we rescale by setting $W_r(s) = W(s/r)$ in order to obtain a finite energy density per unit area. 

The bending terms shall be given in terms of linear ``cosine-type dihedral springs'', which associate energy to the cosine of the dihedral angle between two adjacent lattice cells. More precisely, we set 
\begin{align}\label{eq:E-bend}
\begin{split}
  E^{\rm bend}_r(f) 
  &= D \sum_{\triangle,\triangle' \in {\cal C}_r \atop \text{neighbors}} 
     \left(1 - \cos\widehat{\triangle\triangle'} \right) \\ 
  &= \frac{D}{2} \sum_{\triangle,\triangle' \in {\cal C}_r \atop \text{neighbors}} |n_{\triangle} - n_{\triangle'}|^2, 
\end{split}
\end{align}
where ${\cal C}_r$ denotes the set of equilateral triangles $\triangle \subset U$ of sidelength $r$ with vertices in ${\cal L}_r$;
$\widehat{\triangle \triangle'}$ is the dihedral angle between $f(\triangle)$ and $f(\triangle')$; 
$n_{\triangle}$ is the unit normal to $f(\triangle)$;
and the summations run over all the pairs $\triangle,\triangle' \in {\cal C}_r$  which share a common side. (Note that we tacitly assume that all $f(\triangle)$ are non-degenerate, the derivative $Df$ has rank 2 everywhere and the normals are chosen with consistent orientation.)

We will now consider the limiting energy expressions as $r \to 0$ separately. 

\section{Limiting stretching energy}\label{stretching}

The computation of the continuum limit $r \to 0$ of an energy of the form \eqref{eq:E-stretch} is not new and rather straightforward. We include the short argument for the sake of completeness. 

\subsection{Finite elasticity regime}

Denote by $(Df)_{\triangle}$ the average value of $Df$ on $\triangle \in {\cal C}_r$ and introduce the lattice vectors 
$$ \lat_1 = {-1/2 \choose \sqrt{3}/2}, \quad 
 \lat_2 = {-1/2 \choose -\sqrt{3}/2}, \quad 
 \lat_3 = {1 \choose 0}. $$ 
Then \eqref{eq:E-stretch} can be rewritten as 
\begin{align*}
E^{\rm stretch}_r(f) = \frac{1}{2} \sum_{\triangle \in {\cal C}_r} \sum_{i = 1}^3 r^2 W(|(Df)_{\triangle} \lat_i|) + O(r),  
\end{align*} 
where the error term $O(r)$ accounts for both the linearization error and possible neglection of springs near the boundary. Since $|\triangle| = \sqrt{3}r^2/4$, a Riemann sum argument now gives 
\begin{align*}
  E^{\rm stretch}(f) 
  &:= \lim_{r \to 0} E^{\rm stretch}_r(f)  \\
  &= \frac{2}{\sqrt{3}} \sum_{i = 1}^3 \int_U W(|Df(u) \, \lat_i|) \, du   
\end{align*} 
or, in terms of the first fundamental form $\I(u) = (Df(u))^T Df(u)$, i.e., the right Cauchy-Green strain tensor, of the surface 
\begin{align*}
E^{\rm stretch}(f)  
= \frac{2}{\sqrt{3}} \sum_{i = 1}^3 \int_U W(\sqrt{\langle \lat_i, \I(u) \, \lat_i \rangle}) \, du,    
\end{align*} 
$\langle \cdot, \cdot \rangle$ denoting the Euclidean scalar product. In coordinates $\I(u) = (g_{ij})$ this expression becomes 
\begin{align*}
E^{\rm stretch}(f)  
&= \frac{2}{\sqrt{3}} \int_U \bigg( 
   W\bigg(\frac{\sqrt{g_{11} - 2 \sqrt{3} g_{12} + 3 g_{22}}}{2}\bigg) \\ 
&\qquad\qquad\qquad   + W\bigg(\frac{\sqrt{g_{11} + 2 \sqrt{3} g_{12} + 3 g_{22}}}{2}\bigg) 
   +  W(\sqrt{g_{11}}) \bigg) \, du. 
\end{align*} 
Note that our derivation gives the explicit error bound 
$$ |E^{\rm stretch}_r(f) - E^{\rm stretch}(f)| = O(r) $$ 
and it is not hard to see that this bound is optimal.

\subsection{Infinitesimal elasticity regime}

Here the rescaled Green-St.\ Venant strain tensor $\E_r = \frac{(Df)^T Df - \Id}{2 r}$ is of order one and so $f$ is an almost isometric immersion. More precisely, we assume that $f = f_r$ are lattice defomations such that $\E = \lim_{r\to 0} \E_r$ exists. The contributions $W(|(Df)_{\triangle} \lat_i|)$ to the energy can be Taylor expanded as $W(|(Df)_{\triangle} \lat_i|) = W(1 + r \langle \lat_i, (\E_r)_{\triangle} \lat_i \rangle + O(r^2)) = \frac{1}{2} W''(1) r^2 \langle \lat_i, (\E_r)_{\triangle} \lat_i \rangle^2 + O(r^3)$ and we obtain from \eqref{eq:E-stretch} that 
\begin{align*}
  \lim_{r \to 0} E^{\rm stretch}_r(f_r) 
  &= \lim_{r \to 0} \frac{1}{4} \sum_{\triangle \in {\cal C}_r} \sum_{i = 1}^3 
     W''(1) r^2 \langle \lat_i, (\E_r)_{\triangle} \lat_i \rangle^2 + O(r) \\ 
  &= \int_U \mu \trace(\E^2) + \frac{\lambda}{2} (\trace \E)^2 \, du 
\end{align*} 
is isotropic with Lam{\'e} constants $\mu = \lambda = \frac{\sqrt{3} W''(1)}{4}$. 
Such a result  is in agreement with the predictions given in \citeauthor{Seung:1988}, \citeyear{Seung:1988}; \citeauthor{Disher:1997}, \citeyear{Disher:1997}; \citeauthor{Zhou:1997}, \citeyear{Zhou:1997}. Here the optimal error bound follows as $O(r + |\E_r - \E|)$.

\section{Limiting bending energy}\label{bending}

Our derivation of the limiting bending energy is at variance with earlier computations (see \citeauthor{Seung:1988}, \citeyear{Seung:1988}; \citeauthor{Gompper:1996}, \citeyear{Gompper:1996}; \citeauthor{Disher:1997}, \citeyear{Disher:1997}), and relies on the first and second fundamental forms of $f(U)$. 
We define the nearest neighbor vectors of the dual lattice as 
$$ \latt_1 = {1/2 \choose 1/2\sqrt{3}}, \quad 
 \latt_2 = {-1/2 \choose 1/2\sqrt{3}}, \quad 
 \latt_3 = {0 \choose -1/\sqrt{3}}. $$ 

Rewriting \eqref{eq:E-bend} as 
\begin{align*}
E^{\rm bend}_r(f) 
= \frac{D}{4} \sum_{\triangle \in {\cal C}_r} \sum_{\triangle' \in {\cal C}_r \atop \text{neighbor of } \triangle} |n_{\triangle} - n_{\triangle'}|^2, 
\end{align*} 
we are led to compute the individual contributions $|n_{\triangle} - n_{\triangle'}|^2$ for a fixed triangle $\triangle$. After a rigid transformation, which in particular leaves the first fundamental form $\I(u) = (Df(u))^T Df(u)$ and the second fundamental form 
$$ \II(u) 
 = \sum_{i=1}^3 n_i(u) D^2 f_i(u) $$ 
($n(u) = \frac{\partial_{u_1} f(u) \times \partial_{u_2} f(u)}{|\partial_{u_1} f(u) \times \partial_{u_2} f(u)|}$ the unit normal) invariant, we may assume that the barycenter of $\triangle$ is $0$ and $f = (f_1, f_2, f_3)$ is such that $f(0) = 0$ and $Df_3 = 0$. (See, e.g., \citeauthor{Kuehnel}, \citeyear{Kuehnel} for a basic introduction to the differential geometry of surfaces.)

After a tedious but straightforward computation, using that 
\begin{align*}
  \I(0) 
  &= (D(f_1, f_2))^T D(f_1, f_2)(0), \\ 
  n(0) 
  &= (0,0,1) \quad \mbox{ and so } \\ 
  \II(0) 
  &= D^2 f_3(0), 
\end{align*}
one finds 
\begin{align*}
&\sum_{\triangle' \in {\cal C}_r \atop \text{neighbor of } \triangle} |n_{\triangle} - n_{\triangle'}|^2 \\ 
&= \frac{r^2}{12 \det \I(0)} \sum_{i = 1}^3 \langle \lat_i \I(0) \, \lat_i \rangle 
  \left( 12 \langle \latt_i, \II(0) \, \latt_i \rangle - \trace \II(0) \right)^2 + O(r^3).  
\end{align*} 
A Riemann sum argument finally yields 
\begin{align*}
  E^{\rm bend}(f) 
  &:= \lim_{r \to 0} E^{\rm bend}_r(f) \\ 
  &= \int_U \frac{D}{12 \sqrt{3}} \sum_{i = 1}^3 
  \frac{\langle \lat_i, \I \, \lat_i \rangle 
  \left( 12 \langle \latt_i, \II \, \latt_i \rangle - \trace \II \right)^2}{\det \I} \, du,
\end{align*} 
which in coordinates $\I(u) = (g_{ij})$ and $\II(u) = (h_{ij})$ can also be written as 
\begin{align}\label{Ebend:finite}
\begin{split}
E^{\rm bend}(f) 
= &\int_U \frac{D}{4 \sqrt{3} \det (g_{ij})} ({g_{11} (h_{11}^2 + 2 h_{12}^2 - 2 h_{11} h_{22} + 3 h_{22}^2)} \\
&  \ \ \ \ \ \ \ \ \ \ \ \ \ \ \ \ \ \ \ \ \ \ \ \ {-8 g_{12} h_{11} h_{12} + 2 g_{22} (h_{11}^2 + 3 h_{12}^2)}) \, du. 
\end{split}
\end{align} 
Again our derivation yields the optimal error estimates 
\begin{align*}
  |E^{\rm bend}(f) - E^{\rm bend}(f)| = O(r). 
\end{align*}

The bending energy density is quadratic in $\II$ and can be expressed as 
\begin{align*}
\frac{D}{4 \sqrt{3} \det \I} \langle \II, \mathbb{D} \, \II \rangle 
= \frac{D}{4 \sqrt{3} \det \I} \sum_{i,j,k,l = 1}^2 d_{ijkl} h_{ij} h_{kl}
\end{align*}
in terms of the fourth order tensor $\mathbb{D}= (d_{ijkl})$ with $d_{ijkl} = d_{klij}$, $d_{ijkl} = d_{jikl}$ and 
\begin{align*} 
& d_{1111} = g_{11} + 2 g_{22},
& & d_{1212} = g_{11} + 3 g_{22}, 
& & d_{2222} = 3 g_{11}, \\ 
& d_{1112} = -2 g_{12}, 
& & d_{1122} = -g_{11}, 
& & d_{1222} = 0. 
\end{align*}
This in turn reduces to $\langle \h, {\D} \h \rangle$ with $\h = (h_{11}, h_{22},2 h_{12})$ and the second order $3 \times 3$ tensor 
\begin{align}
& {\D} 
= \frac{D}{2 \sqrt{3} \det (g_{ij})}  
  \begin{pmatrix} g_{11} + 2 g_{22}  & -g_{11} & - 2g_{12} \\
                            -g_{11} & 3 g_{11} & 0 \\ 
                  -2 g_{12} & 0 & \frac{1}{2} (g_{11} + 3 g_{22} )  
  \end{pmatrix}, 
  \label{Dmatrix}
\end{align}
from which the nominal bending stresses and the the referential bending moduli can be readily read off by differentiation with respect to $\h$. Note carefully that these will depend on the first fundamental form $(g_{ij})$ thus introducing a coupling between stretching and bending contributions.
Symmetrically, the differentiation of the bending energy density with respect to $g_{ij}$ provides contributions to the nominal stretching stresses and referential stretching moduli. 
We remark that (\ref{Ebend:finite}) accounts for possible heavy stretching and shearing deformations of the reference lattice, due to the presence of the first fundamental form $\I(u)$ in the limiting energy. We devote the following section to the derivation of the complete set of elastic moduli that characterize the present network model.

For isometric immersions, when $\I(u) = \Id$,  formula  (\ref{Ebend:finite}) reduces to 
\begin{align}\label{eq:Helfrich}
E^{\rm bend}(f) 
= \frac{D}{4 \sqrt{3}} \int_U (3 H^2 - 8 K) \, du, 
\end{align} 
where $H = \trace{\II}$ is twice the mean curvature and $K = \det \II$ is the Gauss curvature of the surface, which provides a good approximation to the energy also when $\I \approx \Id$ (\textit{bending-dominated} deformation regime). It has to be noted that $\I \equiv \Id$ would imply that $f$ is an isometric immersion and thus $K \equiv 0$. 
However, general surfaces can be triangulated with $\I \approx \Id$ if one allows for defects in the reference triangular lattice on a mesoscopic scale. More precisely: Choosing a partition of the surface into regions of area and boundary length scaling with $\tilde{r}^2$ resp.\ $\tilde{r}$ for $r \ll \tilde{r} \ll 1$ one can parameterize the surface with almost trivial first fundamental form if one allows for defects in the reference lattice along the mesoscopic boundary of these elements. The number of atomic bonds across this boundary is $O((\tilde{r} r)^{-1})$, which is much less than the total number of bonds, which scales with $r^{-2}$. In the limit $r,\tilde{r} \to 0$ the contribution of the defects becomes negligible. Also the stretching terms will vanish as $\I$ approaches $\Id$, and so the limiting continuum energy is given by \eqref{eq:Helfrich}. More precisely, the extra stretching energy from the lattice defects is bounded by $O(\frac{r}{\tilde{r}})$, while $|\I - \Id| = O(\tilde{r})$, leading to $O(\tilde{r}^2)$ contributions to $E^{\rm stretch}$ and, in general, $O(\tilde{r})$ contributions to $E^{\rm bend}$. The optimal error estimate taking into account all of these contributions is obtained by choosing the scaling $\tilde{r} \sim \sqrt{r}$, which leads to an overall error of order $O(\sqrt{r}$). 

As already observed, formula (\ref{eq:Helfrich}) predicts bending modulus
\begin{align}
\kappa_H   & = 
\frac{\sqrt{3} D}{2}
\label{kappaHlin}
\end{align} 
for any geometry, and Gaussian rigidity $\kappa_G=-4\kappa_H/3$, differently from what found in \citeauthor{Seung:1988}, \citeyear{Seung:1988} and \citeauthor{Gompper:1996}, \citeyear{Gompper:1996}, and in agreement with the estimates of $\kappa_H$ and $\kappa_G$ given in \citeauthor{Lidmar:2003}, \citeyear{Lidmar:2003}.

\section{Continuum model and elastic moduli}\label{sec:moduli}

Let us introduce the total limiting energy density   
\begin{align*} 
W^{\rm limit}(\bI(u),\bI\bI(u)) \ =  \ W^{\rm stretch}(\bI(u)) \ + \ W^{\rm bend}(\bI(u),\bI\bI(u))
\end{align*} 
where in the regime of finite elasticity with linear springs, i.e., $W(s) = \frac{k}{2}(s - 1)^2$,
\begin{align*} 
& W^{\rm stretch}(\bI(u)) \ = \  
\frac{k}{2 \sqrt{3}} 
   \bigg( 3 g_{11} + 3 g_{22} + 6 - 2 \sqrt{g_{11} - 2 \sqrt{3} g_{12} + 3 g_{22}}  \\ 
& \qquad\qquad\qquad\qquad\qquad\qquad\qquad - 2 \sqrt{g_{11} + 2 \sqrt{3} g_{12} + 3 g_{22}} 
   - 4 \sqrt{g_{11}} \bigg), 
\\
& W^{\rm bend}(\bI(u),\bI\bI(u) )   \ = \ \frac{\sqrt{3} D}{12 (g_{11} g_{22} - g_{12}^2)} \ ({g_{11} (h_{11}^2 + 2 h_{12}^2 - 2 h_{11} h_{22} + 3 h_{22}^2)} \nn \\
& \qquad\qquad\qquad\qquad\qquad\qquad\qquad\qquad\qquad {-8 g_{12} h_{11} h_{12} + 2 g_{22} (h_{11}^2 + 3 h_{12}^2)}). 
\end{align*} 

By differentiating $W^{\rm limit}$ with respect to $\bg = (g_{11},g_{22},2g_{12})$ and $\bh$, we get the following expressions for the referential stretching stresses ${ \bN}$; bending stresses ${ \bM}$;  stretching moduli ${ \bA}$; 
and  bending-stretching moduli ${ \bB}$ 
\begin{align*} 
{ \bN} = 2 \frac{\partial W^{\rm limit}}{\partial \bg}, \ \ \
{ \bM} = \frac{\partial W^{\rm limit}}{\partial \bh}
\end{align*} 
\begin{align*} 
{ \bA} = 4 \frac{\partial^2 W^{\rm limit}}{\partial \bg^2}, \ \ \
{ \bB} = 2 \frac{\partial^2 W^{\rm limit}}{\partial \bg \partial \bh}, \ \ \
{ \bD} = \frac{\partial^2 W^{\rm limit}}{\partial \bh^2}
\end{align*} 
which define the limiting elastic response of the network, in association with the referential bending moduli (\ref{Dmatrix}). The overall (stretching-bending) elasticity matrix consists of the second order $6 \times 6$ tensor 
\begin{align} 
{\bC}=\left[
\begin{array}{c|c}
{ \bA} & { \bB} \\
\hline 
{ \bB}^T & { \bD}
\end{array}\right]
\label{GlobCmat}
\end{align} 
We provide in an Appendix the expressions of the arrays $\bN, \ \bM, \bA, \ \bB$ for the general case of finite elasticity, which corresponds to fully anisotropic response of the network (full rank ${\bC}$, cf.\ also \citeauthor{Disher:1997}, \citeyear{Disher:1997}). 

Let us introduce now the inverse ${\bC}^{-1}=(C_{ij}^{-1})$ of $\bC$, and the `engineering' elastic coefficients
\begin{align} 
& { \kappa}_A  =  \frac{1}{{C}_{11}^{-1}+2{C}_{12}^{-1}+{C}_{22}^{-1}}, \ \
{ G}_{12} \ = \ \frac{1}{{C}_{33}^{-1}},
\label{kappaAGnlin}
\end{align} 
\begin{align}
\kappa_H   & = 
\ \frac{C_{44} + C_{55} +2 C_{45} + C_{66}}{4} \ = \
\frac{\sqrt{3} D \left( g_{11} + 2 g_{22} \right) }{6 \det \bI}
\label{kappaHnlin}
\end{align} 
representing the  \textit{area compression modulus}, \textit{in-plane shear modulus} and \textit{bending modulus} of the limiting surface, respectively. The definitions (\ref{kappaAGnlin})  are consistent with the notation usually employed for the in-plane engineering constants of anisotropic plates (refer to standard textbooks on the mechanics of composite materials, such as, e.g., \citeauthor{Nettles1994}, \citeyear{Nettles1994}; Sec.\ V.C). For what instead concerns the bending modulus (or rigidity) $\kappa_H$, we recall that the theory of curvature elasticity of fluid membranes defines such a quantity as the
rigidity associated with the sum of the principal curvatures, assuming isotropic bending response (\citeauthor{Helfrich:1973}, \citeyear{Helfrich:1973}). Within the present anisotropic framework, it is easily verified that (\ref{kappaHnlin}) conventionally defines  $\kappa_H$ as the arithmetic mean of the bending rigidities to the incremental (virtual) strains with $\bg'=\bg''=\mbs{0}$; $\bh'=\{1/2,1/2,1\}$ and $\bh''=\{1/2,1/2,-1\}$ from the current configuration (noticeable strain rates that modify the trace $H = \trace{\II}=h_1+ h_2$ of the curvature tensor, leaving the Gaussian curvature $K = \det \II=h_1 h_2 - (h_3/2)^2$ and the first fundamental forms unchanged).    
We wish again to remark that the elasticity matrix (\ref{GlobCmat}) depends both on the first and second fundamental forms of the current (deformed) configuration of the limiting surface (cf.\ the formulae given in the Appendix), introducing a type of stretching-bending coupling that is not accounted for in standard energy functionals of fluid membranes (refer e.g. to  \citeauthor{Lipowski:1990}, \citeyear{Lipowski:1990}; \citeauthor{HelfrichKozlov:1993}, \citeyear{HelfrichKozlov:1993}).

We now consider the bending-dominant regime with $\I \approx \Id$. It is not difficult  to verify that, in such a regime, equations (\ref{kappaAGnlin}) reduce to
\begin{align}
& { \kappa}_A  = \frac{\sqrt{3} k}{2}
\label{kappaAbend}
\end{align} 
\begin{align}
\begin{split} 
  { G}_{12} 
& =   \frac{1}{16 \sqrt{3} D (h_{11}^2 + 3 h_{12}^2 ) + 12 \sqrt{3}  k}
\  (8 D^2 \ (h_{11}^4 + 6 h_{12}^4 \\
&\quad - 2 h_{11}^3 h_{22}  - 6 h_{11}h_{12}^2h_{22} + 9 h_{12}^2h_{22}^2 - 3 h_{11}^2 (h_{12}^2-h_{22}^2) )
 \\
& \quad + \ 6 D k \ (3 h_{11}^2 + 8 h_{12}^2 - 2h_{11}h_{22} + 3 h_{22}^2)  \ + \ 9 k^2) 
\end{split}
\label{G12bend}
\end{align}

Formula (\ref{kappaAbend}) is in agreement with the result presented in \citeauthor{Zhou:1997} (\citeyear{Zhou:1997}), \citeauthor{Seung:1988} (\citeyear{Seung:1988}), \citeauthor{Disher:1997} (\citeyear{Disher:1997}) for the area compression modulus of triangular nets in the small strain regime. On the contrary, formula (\ref{G12bend}) differs from the analogous one provided in the same works for the in-plane shear modulus in the regime under examination ($\mu={\sqrt{3} k}/{4}$, assuming pure planar deformation), due to the presence of curvature terms. The presence of such terms in the expression of $G_{12}$ implies that the shear response of the examined membrane network is actually anisotropic and curvature dependent, when marked bending deformation occurs. In the fully infinitesimal deformation regime ($\I \approx \Id$ and $D |\bh |^2 \ll k$) we instead recover the result $\mu \equiv G_{12}={\sqrt{3} k}/{4}$ from (\ref{G12bend}). It is also immediate to verify that formula (\ref{kappaHnlin}) reduces to (\ref{kappaHlin}) when $g_{11}\approx g_{22} \approx 1$.

\section{Special cases}\label{cases}
We examine in the present section the two special cases of  cylindrical and spherical networks covered with $N_{\triangle}$ equilateral triangles.
In both such cases, we show that formula (\ref{eq:Helfrich}) exactly reproduces the limit value of the total dihedral energy of the network for $N_{\triangle}\to\infty$. The given examples, which allow for a comparison with explicit formulae in the literature, are aimed to analytically illustrate the universal nature of formula (\ref{eq:Helfrich}), which predicts equal bending stiffnesses for any geometry, namely $\kappa_H= \sqrt{3} D /2$ and $\kappa_G=-4\kappa_H/3=-2\sqrt{3}D/3$, at variance with the prediction $\kappa_H=-\kappa_G=\sqrt{3}D/2$ of \citeauthor{Seung:1988} (\citeyear{Seung:1988}). We also provide the expressions of the in-plane shear moduli of the same networks accounting for finite curvature. The area compression modulus is given by equation (\ref{kappaAbend}) in the regime under consideration, for any geometry.

	\subsection{Cylindrical network}	
\citeauthor{Gompper:1996} (\citeyear{Gompper:1996}) compute the following value of the dihedral energy per unit of length of an infinitely long cylinder of unit radius covered with $N_{\triangle}$ equilateral triangles (again per unit of length)
\begin{align*}
E^{\rm dihedral}_{\rm cylinder}
=\frac{\sqrt{3}\pi D}{2}\cdot \left ( 1-\frac{\pi}{4\sqrt{3}N_{\triangle}}+O\left (1/N^2_{\triangle}\right ) \right )
\to \frac{\sqrt{3}\pi D}{2},
\end{align*} 
which reduces to \eqref{eq:Helfrich} with $H=1, K=0$ and $|U| = 2 \pi$. Examining now the in-plane shear modulus, and considering a cylinder of finite radius $R$, from (\ref{G12bend}) we deduce the result 
\begin{align*} 
&  \mu \ \equiv \ { G}_{12}  \ = \ \frac{\sqrt{3} k}{4} \ + \ \frac{\sqrt{3} D H^2}{6} \ = \  \frac{\sqrt{3} k}{4} \ + \ \frac{\sqrt{3} D}{6 R^2}.
\end{align*}

	\subsection{Spherical network}	
The total dihedral energy of a spherical network covered with $N_{\triangle}$ equilateral triangles is given by formula (19) of \citeauthor{Gompper:1996} (\citeyear{Gompper:1996}):

\begin{align*}
E^{\rm dihedral}_{\rm sphere}
=\frac{4\pi D}{\sqrt{3}}\left ( 1+\frac{2\pi}{3\sqrt{3}N_{\triangle}}+O\left ( 1/N_{\triangle}^2 \right ) \right )
\to \frac{4\pi D}{\sqrt{3}}.
\end{align*} 
This exactly coincides with \eqref{eq:Helfrich} for the sphere with $H=2/R, K=1/R^2$ and $|U| = 4 \pi R^2$. For what concerns the in-plane shear modulus of a spherical network, from (\ref{G12bend}) we easily deduce 

\begin{align} 
&  \mu \ \equiv \  { G}_{12}  \ = \    \frac{\sqrt{3} k}{4} \ + \ \frac{\sqrt{3} D H^2}{12}
\ = \    \frac{\sqrt{3} k}{4} \ + \ \frac{\sqrt{3} D}{3 R^2}.
\label{G12sphere}
\end{align}

\section{Experimental characterization of material constants}\label{expvalidation}

\noindent Thinking of cell membranes, and especially of blood cell membranes, the results presented in the previous sections are useful to conduct an experimental characterization of the material constants $k$ and $D$ through methods like micropipette aspiration, optical tweezes and fast phase constrast microscopy, among others (refer, e.g., to \citeauthor{BookMofradKamm:2006} (\citeyear{BookMofradKamm:2006}), Chapt.\ 2). 

\begin{figure}[htbp]
  \begin{center}
    \epsfig{file=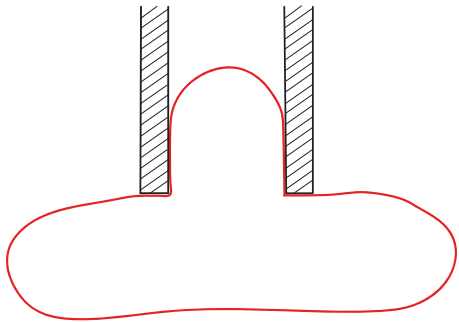,angle=0,width=60mm} 
   \caption{Aspiration of a red blood cell into a micropipette.}
    \label{micropipette}
  \end{center}
\end{figure}

In particular, micropipette aspiration tests are often used to determine the elastic properties of nonadherent cells, through measurements of  the cortical tension and increase in the surface area (area compression modulus, cf.\ \citeauthor{Evans:1976}, \citeyear{Evans:1976};  \citeauthor{Evans:1977}, \citeyear{Evans:1977});  the length of the cell tongue aspirated into the pipette at increasing values of applied pressure (shear modulus, cf.\ \citeauthor{Waugh:1979}, \citeyear{Waugh:1979}); the critical aspiration pressure at which buckling of aspirated cell tongue occurs (bending modulus, cf.\ \citeauthor{Evans:1983}, \citeyear{Evans:1983}); and other different quantities (refer to the above mentioned works and therein references).
Two measurements are sufficient to determine $k$ and $D$ (typically the area compression modulus and/or the shear modulus, plus the bending modulus), while a third one can be useful for a cross-check.
Fig.\ \ref{micropipette} illustrates the aspiration of a red blood cell into a micropipette. 
When the size $L$ of the aspirated cell tongue is such that this portion of the cell assumes a spherical shape (Fig. \ref{micropipette}), and contemporarly stretching strains are sufficiently small, one can predict the shear modulus through (\ref{G12sphere}), with $R \approx L$. When, instead, the curvature of such an element is associated with large stretching strains, one should numerically predict the elastic moduli through equations (\ref{kappaAGnlin}) and  (\ref{kappaHnlin}), making use of suitable measurements of the stretching strains.

\section{Conclusions}\label{conclusions}

We have developed universal formulae for the limiting  elastic energy of static membrane networks, which arise in the limit for the number of vertices/particles tending to infinity. The given formulae account for arbitrary geometries, large deformations and bending-stretching coupling. Their particularization to the infinitesimal elastic regime has been carried out, and comparisons with results available in the literature have been established (\citeauthor{Seung:1988}, \citeyear{Seung:1988}; \citeauthor{Gompper:1996}, \citeyear{Gompper:1996}; \citeauthor{Disher:1997}, \citeyear{Disher:1997}; \citeauthor{Zhou:1997}, \citeyear{Zhou:1997}; \citeauthor{Lidmar:2003}, \citeyear{Lidmar:2003}). It has been shown that the formulae here presented overcome some deficiencies of earlier results for triangulated networks endowed with cosine-type dihedral angle potentials, which are related to shape-dependence of the bending stiffness and sensitivity to mesh distortion of the limiting bending energy (\citeauthor{Muller:2006}, \citeyear{Muller:2006}; \citeauthor{Gompper:1996}, \citeyear{Gompper:1996}; \citeauthor{Nelson:2004}, \citeyear{Nelson:2004}; \citeauthor{Espriu:1987}, \citeyear{Espriu:1987}; \citeauthor{Baillie:1990}, \citeyear{Baillie:1990}). We remark that the finite-elasticity version (\ref{Ebend:finite}) of the present limiting bending energy is able to handle heavily distorted triangulations, due to the presence of finite membrane deformation terms. We have derived the entire set of the elastic coefficients of the examined network model, obtaining general expressions of such quantities, which depend on the first and second fundamental forms $g_{ij}$ and $h_{ij}$ of the limiting surface (full bending-stretching coupling). We have also obtained simplified expressions of the {area compression modulus}, {in-plane shear modulus} and {bending modulus} in the bending-dominant regime, which are useful to carry out the experimental identification of such quantities. 
We address to future work the derivation of deformation-dependent elastic moduli of fluctuating membrane networks undergoing large stretching and/or bending deformations, on combining the formulae for the limiting energies and elastic moduli here proposed with MD or Monte Carlo simulations, and the local maximum-entropy approximation scheme presented in (\citeauthor{biocurvature}, \citeyear{biocurvature}); an extension of the study of phase transition phenomena for Hookean spring networks presented in \citeauthor{Disher:1997} (\citeyear{Disher:1997}), in order to account for the coupling of in-plane and out-of-plane deformations in the finite elasticity regime; 
as well as the experimental characterization of the examined model through established laboratory tests, and innovative techniques to be designed via computer simulation.

\section*{Acknowledgements}
Bernd Schmidt thanks the Department of Civil Engineering of the University of Salerno, where part of this work has been completed, for its hospitality. Fernando Fraternali gratefully acknowledges the precious advices received by Chris Lorenz and Gianluca Marcelli (Division of Engineering, King's College London) about the physical aspects of the present work.

\section*{Appendix. Supplementary data}\label{Appendix}
The streching and bending stresses and moduli from Sect.\ \ref{sec:moduli} can be easily calculated with a computer algebra program. Below we give a {\tt mathematica} code to compute these quantities. A complete list of these constants can be found as supplementary data associated with this article in the online version.

Introduce the energy density as: 
\begin{tt}
\begin{footnotesize}
\begin{quote}
\noindent 
WlimG[\{g1\underline{~}, g2\underline{~}, g3\underline{~}, h1\underline{~}, h2\underline{~}, h3\underline{~}\}] \\ 
     = (k / (2 Sqrt[3])) \\ 
     * ( 3 g1 + 3 g2 + 6 - 2 Sqrt[g1 - 2 Sqrt[3] g3 /2 + 3 g2] - \\  
     2 Sqrt[g1 + 2 Sqrt[3] g3/2 + 3 g2] - 4 Sqrt[g1]) \\ 
     + ((Sqrt[3] D) / (12 (g1 g2 - g3$\hat{~}$2/4)) \\ 
     * ( g1 (h1$\hat{~}$2 + 2 h3$\hat{~}$2/4 - 2 h1 h2 + 3 h2$\hat{~}$2) - 8 g3 h1 h3/4 + \\ 
      2 g2 (h1$\hat{~}$2 + 3 h3$\hat{~}$2/4) )); \\ 
WlimE[\{e1\underline{~}, e2\underline{~}, e3\underline{~}, h1\underline{~}, h2\underline{~}, h3\underline{~}\}] \\ 
:= Wlim[\{2 + 2 e1, 2 + 2 e2, 2 e3, h1, h2, h3\}] ;
\end{quote}
\end{footnotesize}
\end{tt}

The 6 dimensional vector of the stretching and bending stresses $({\bf N},  {\bf M})$ is then given by:

\begin{tt}
\begin{footnotesize}
\begin{quote}
\noindent 
D[WlimE[\{e1, e2, e3, h1, h2, h3\}], \{\{e1, e2, e3, h1, h2, h3\}, 1\}] \\ 
/. \{e1 $\to$ 1 - g11/2, e2 $\to$ 1 - g22/2, e3 $\to$ -g12, \\ 
h1 $\to$ h11, h2 $\to$ h22, h3 $\to$ 2 h12\} // MatrixForm // Simplify
\end{quote}
\end{footnotesize}
\end{tt}

The $6 \times 6$ stretching- bending elasticity matrix ${\bf C}$ from \eqref{GlobCmat} is given by:

\begin{tt}
\begin{footnotesize}
\begin{quote}
D[WlimE[\{e1, e2, e3, h1, h2, h3\}], \{\{e1, e2, e3, h1, h2, h3\}, 2\}] \\ 
/. \{e1 $\to$ 1 - g11/2, e2 $\to$ 1 - g22/2, e3 $\to$ -g12, \\ 
h1 $\to$ h11, h2 $\to$ h22, h3 $\to$ 2 h12\} // MatrixForm // Simplify
\end{quote}
\end{footnotesize}
\end{tt}

\typeout{References}

\end{document}